\documentclass[doublecol]{epl2} 

\usepackage{amsmath}
\usepackage{amssymb}

\def \ba\begin{align}
\def \ea\end{align}

\def \S{{\mathcal{S}}}

\usepackage{graphicx}
\usepackage{dcolumn}
\usepackage{bm}
\usepackage{color}

\usepackage{array}
\newcolumntype{x}[1]{>{\centering\arraybackslash\hspace{0pt}}p{#1}}

\def \emanuele#1{{\color{black}#1\color{black}}}

\begin{document}


\title{\emanuele{Complete characterization of spin chains with two Ising symmetries}}

\author{Bat-el Friedman\inst{1,*}, Atanu Rajak\inst{1,*}, Emanuele G. Dalla Torre\inst{1}}
\institute{
\inst{1}Department of Physics and Center for Quantum Entanglement Science and Technology, Bar-Ilan University, 52900 Ramat Gan, Israel\\
$^*$B.~F. and A.~R. equally contributed to this work\\
}
\shortauthor{B.-e.Friedman, A. Rajak, {\it et al.}}

\pacs{03.65.Vf}{Topological phases (quantum mechanics)}
\pacs{05.30.Rt}{quantum phase transitions}
\pacs{75.10.Pq}{spin chain models}





\abstract{Spin chains with two Ising symmetries are the Jordan-Wigner duals of one-dimensional interacting fermions with particle-hole and time-reversal symmetry. From earlier works on Majorana chains, it is known that this class of models has 10 distinct topological phases. In this paper, we analyze the physical properties of the correspondent 10 phases of the spin model. In particular, thanks to a set of two non-commuting dualities, we determine the local and non-local order parameters of the phases. We find that 4 phases are topologically protected by the Ising symmetries, while the other 6 break at least one symmetry. Our study highlights the non-trivial relation between the topological classifications of interacting bosons and fermions.}

%

\maketitle


\def \mysection#1{{\it\bf #1 --}}

\mysection{\label{intro} Introduction}
Conventional phases of matter, like in the Ising ferromagnet, spontaneously break a symmetry and show a non-vanishing local order parameter. Phases that do not break any symmetry are called topological, and include the quantum Hall phases \cite{ando1975theory,klitzing1980new,laughlin1981quantized,halperin1982quantized}, the topological insulators \cite{kitaev2009periodic,ryu2010topological,hasan2010colloquium}, and the Haldane phase \cite{haldane1983continuum,haldane1983nonlinear,affleck1987rigorous, dalla2006hidden,auerbach2012interacting}. In the last decade, these phases were classified and generalized within the framework of symmetry protected topological (SPT) phases \cite{turner2013beyond,chen2013symmetry,senthil2015symmetry}. In particular, for interacting bosons, a complete classification of SPT phases was achieved by observing that, for a given symmetry group $G$, the number of distinct topological phases is equal to the size of a cohomology group of $G$ \cite{chen2011classification,chen2011complete,chen2012symmetry,chen2013symmetry}. In analogy to the classification of all crystal structures, the key question now is to associate specific materials or Hamiltonians to the different SPT phases, and to study their physical properties. Here we address this question for Ising spin chains.



%


The study of the topological phases of spin chains emanated from the detailed study of the Haldane phase. 
It was found that this phase can be protected by several symmetry groups $G$: (i)~spin rotations, (ii)~time reversal, (iii)~lattice inversion, and (iv)~rotations of $\pi$ around three orthogonal axes ($Z_2\times Z_2$) \cite{berg2008rise,pollmann2012symmetry}. The cohomology group of these symmetries is $Z_2$, indicating that these classes of models have only one topologically non-trivial phase, the Haldane phase\cite{pollmann2010entanglement,chen2011classification,schuch2011classifying}.

\begin{table}[ht!]
	\renewcommand{\arraystretch}{1.25}
	\begin{small}
		\begin{tabular}{|x{0.45cm}|x{0.45cm}|x{0.6cm}|x{1.5cm}|x{1.7cm}|x{1.5cm}|}
			\hline
			$I_z$ & $I_y$ & $I_zI_y$ & symmetry group $G$ & cohomology group \cite{chen2013symmetry} & \emanuele{parent Hamilton.}\\
			\hline
			yes & yes & yes & $Z^T_2\times Z_2^T$ & $\mathbb{Z}_2\times \mathbb{Z}_2$ & $m=0,\pm 2, 4$\\
			\hline
			yes & no & no & $Z_2^T$ & $\mathbb{Z}_2$ & $m=1,-3$\\
			\hline
			no & yes & no & $Z_2^T$& $\mathbb{Z}_2$  & $m=-1,3$\\
			\hline
			no & no & yes & $Z_2$ & $\mathbb{Z}_1$ & nematic\\
			\hline
			no & no & no & $Z_1$ & $\mathbb{Z}_1$ &canted\\
			\hline
		\end{tabular}
	\end{small}
	\caption{The 10 phases of one-dimensional (1d) spin models with two Ising symmetries. \emanuele{The first 4 columns show the symmetries preserved by each group of phases and the resulting symmetry group $G$}. The 5$^{th}$ column lists the 2-Borel-cohomology group $\mathcal{H}^{2}[G,U(1)]$, whose number of elements is equal to the number of topologically distinct phases in $1d$ \cite{chen2011complete,chen2012symmetry,chen2013symmetry}. \emanuele{The 6$^{\rm th}$ column lists the parent Hamiltonians, see Eq.~(\ref{eq:H0}).} The 8 phases listed in first 3 rows ($m=-3,...,4$) can be mapped to the 8 phases of interacting fermions with P and T symmetry, see Table \ref{table:order}. The last 2 phases correspond to fermionic phases that spontaneously break T. \label{table2}}
	\vspace{-0.2cm}
\end{table}

A richer topological classification can be obtained by considering models with a symmetry group $Z^T_2\times Z_2^T$, where $Z_2^T$ is an {\it anti-unitary} $Z_2$ symmetry. The correspondent cohomology group is $\mathbb{Z}_2\times \mathbb{Z}_2$, implying the existence of 4 topologically distinct phases that do not break any symmetry \cite{liu2011gapped,chen2012symmetry,chen2013symmetry}. A simple realization of a $Z_2^T\times Z_2^T$ symmetry is offered by spin chains with two Ising symmetries\footnote{The easiest way to see that Ising symmetries are anti-unitary is to note that they flip the sign of the canonical commutation relations of the spins. To better appreciate the distinction between \emanuele{the Ising symmetries of Eq.~(\ref{eq:Ising}) and} the $Z_2\times Z_2$ symmetry associated with $\pi$ rotations around the axes we observe that the term $\sigma^x_{i}\sigma^y_{i+1}\sigma^z_{i+2}$ is \emanuele{symmetric under $\pi$ rotations, but not under the Ising symmetries. And, {\it vice versa}, the term $\sigma^x_{i}$ is symmetric only under the latter}.}
\begin{align} I_z:~\sigma^z\to-\sigma^z,~~{\rm and}~~I_y:~\sigma^y\to-\sigma^y\;.\label{eq:Ising}\end{align}
In addition to the 4 SPT phases, this model has 6 phases that break at least one symmetry, for a total number of 10 phases (see Table \ref{table2}).
In this paper, we characterize the physical properties of these phases by providing exactly diagonalizable parent Hamiltonians, describing their local and nonlocal order parameters, and identifying their edge states. 

\mysection{From non-interacting fermions to spin chains} Our strategy is based on the Jordan-Wigner transformation of known topological phases of 1d fermions. This transformation was first introduced to solve the quantum Ising model, by mapping it to a 1d system of non-interacting fermions~\cite{jordan1928paulische,sachdev2007quantum}. At a critical value of the Ising coupling, the model undergoes a quantum phase transition, where the ground state degeneracy changes from 1 (paramagnet) to 2 (ferromagnet). In the fermionic language, this is due to the appearance of zero-energy Majorana edge modes \cite{kitaev2001unpaired,greiter20141d}. This fermionic system is currently referred to as a Kitaev chain and has become very popular due to its implications for topological quantum computing (see Ref.~\cite{alicea2010non} for an introduction).  

\begin{figure}[t]
\centering
\includegraphics[scale=0.8]{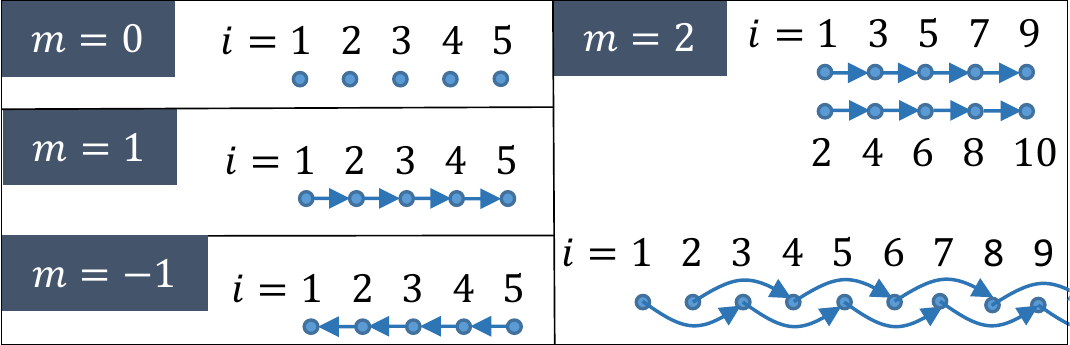}
\caption{Real-space representation of topological phases of non-interacting fermions in 1d, Eq.~(\ref{eq:H0}): the trivial phase ($m=0$) is site factorizable. The phases with $m= \pm 1$ are Kitaev chains with one Majorana mode at each edge. The phases with $|m|>1$ can be  constructed by stacking in parallel several Kitaev chains.  For the phase $m=2$, two equivalent graphical representations are given.\label{fig:scheme2}}
\end{figure}

Further topological phases of 1d fermions can be obtained by stacking in parallel several Kitaev chains \cite{fidkowski2011topological,niu2012majorana}. Reordering the site indexes, this is equivalent to a single chain with longer-range couplings (see Fig.~\ref{fig:scheme2}):
\begin{align}
\label{eq:H0}
H_0^{(m)} = \sum_i & (\psi^\dagger_i\psi_{i+m}+\psi^\dagger_{i+m}\psi_i)\nonumber\\ 
&+ {\rm sign}(m)(\psi^\dagger_i\psi^\dagger_{i+m}+\psi_{i+m}\psi_i)\;.
\end{align} 
Here  the topological index $m \in \mathbb{Z}$ determines the number and flavor of the Majorana modes at each edge \footnote{Majorana fermions appear in two flavors, often denoted by $a$ and $b$: one is the time-reversed conjugate of the other. Note that negative $m$'s correspond to Kitaev chains with inverted site indexes ($i\to-i$).}. Due to the presence of zero-energy Majorana modes, each eigenstate of Eq. ($\ref{eq:H0}$) is $2^{|m|}$-fold degenerate \footnote{The phase $m$ has $2|m|$ Majorana modes ($m$ per side), which can be combined in a set of $|m|$ complex fermions. Note  that in the non-interacting case, each eigenstate has the same degeneracy.}.



%
%
The models~(\ref{eq:H0}) have both time-reversal ($T^2=+1$) and particle-hole ($P^2=+1$) symmetry. Thus, according to the classification of topological phases of non-interacting fermions (topological insulators and superconductors), they belong to the class BDI~\cite{zirnbauer1996riemannian,altland1997nonstandard,kitaev2009periodic,ryu2010topological,hasan2010colloquium}: In 1d, this symmetry group gives rise to $\mathbb{Z}$ distinct topological phases, one for each possible value of $m$.


We now use Eq.~(\ref{eq:H0}) to construct parent Hamiltonians for the topological phases of spin chains. By applying the Jordan-Wigner transformation, one obtains the integrable Hamiltonians: 
\begin{align}
H_0^{(m)}=-\sum_i\left\{
\begin{array}{l l} 
\sigma^z_i \left(\prod_{j=i+1}^{i+m-1}\sigma^x_j\right)\sigma^z_{i+m},&m>0\\
\sigma^x_i,&m=0\\
\sigma^y_i \left(\prod_{j=i+1}^{i+|m|-1}\sigma^x_j\right)\sigma^y_{i+|m|},&m<0
\end{array}\right.\;,
\label{eq:H0spin}
\end{align}
\emanuele{where $\vec{\sigma}_i=(\sigma^x_i,\sigma^y_i,\sigma^z_i)$ are Pauli matrices. These Hamiltonians were first introduced by M. Suzuki \cite{suzuki1971relationship} and are known as generalized XY models, or generalized cluster Ising models \cite{keating2004random,kopp2005criticality,son2011quantum,smacchia2011statistical,niu2012majorana,degottardi2013majorana,lahtinen2015realizing,ohta2016topological,lee2016string,russomanno2017spin}. The $m=2$ model, $H^{(2)}=-\sum_i\sigma^z_i\sigma^x_{i+1}\sigma^z_{i+2}$, is of particular interest for quantum computing: Its ground state, known as the one-dimensional cluster state \cite{briegel2001persistent}, offers the possibility to realize universal one-qubit gates via measurement based quantum computations \cite{raussendorf2001one,doherty2009identifying,stephen2017computational}. The Hamiltonians of the models with $-4<m\leq4$ are explicitly shown in the 2$^{\rm nd}$ column of Table \ref{table:order}. Importantly, the T and P symmetry of the fermionic Hamiltonians imply the two Ising symmetries of Eq.~(\ref{eq:Ising}) for the spin models.}

\mysection{\label{dualities}Two duality transformations} \emanuele{To characterize the phases of the cluster Ising models,} it is useful to introduce two duality transformations~\cite{keating2004random,degottardi2013majorana}. These transformations reflect the symmetry of the model: they map a magnetic field in the $x$ direction to Ising couplings in the $z$ and $y$ direction, respectively. Having introduced the string operator
\begin{align}
\S_i^x=\prod_{j\le i}\sigma_j^x\label{eq:string}\;,
\end{align}
we can formally define the duality transformations 
\begin{align}
D_z:&~\sigma^x_i \to \sigma^z_i\sigma^z_{i+1}~,~~~\sigma^z_i \to \S^x_i\label{eq:Dx}\;;
\end{align}
and
\begin{align}
D_y:&~\sigma^x_i \to \sigma^y_i\sigma^y_{i+1}~,~~~\sigma_i^y\to \S^x_i\;.\label{eq:Dy}
\end{align}
%
\begin{figure}
\centering
\vspace{-0.4cm}
\includegraphics[scale=0.75]{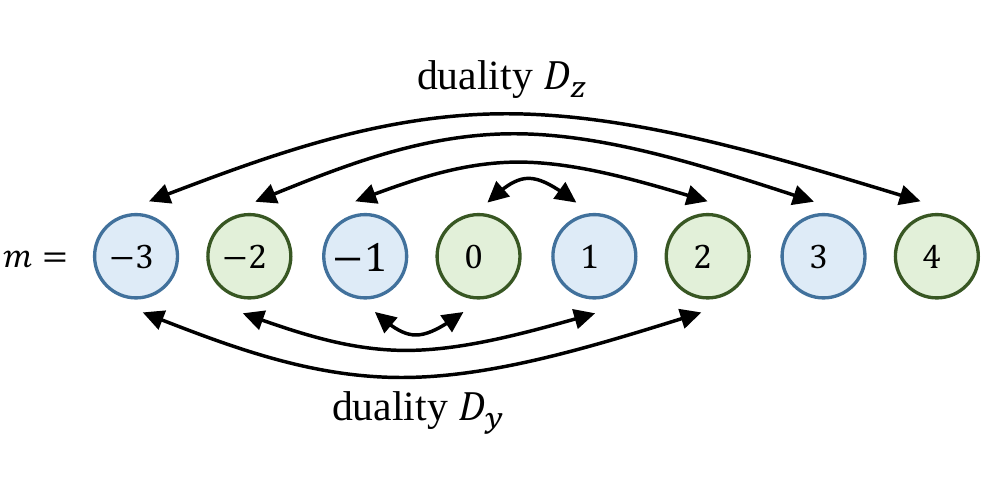}
\vspace{-0.7cm}
\caption{Configuration-space representation of topological phases of spin-1/2 chains, Eq.~(\ref{eq:H0spin}). The phases are connected by the duality transformations $D_z$ and $D_y$, Eqs.~(\ref{eq:H0dualityX}) and (\ref{eq:H0dualityY}). The odd-$m$ phases (green) are characterized by a local order parameter, and the even-$m$ phases (light-blue) by a nonlocal order parameter -- see Table \ref{table:order} for more details. The 8 phases shown in this graph cannot be adiabatically connected even in the presence of integrability-breaking terms.\label{fig:scheme1}}
\end{figure}
By definition, $D_z$ and $D_y$ map $H^{(0)}$, respectively, to $H^{(1)}$ and $H^{(-1)}$. What happens if we now apply $D_z$ to $H^{(-1)}$? By using the identity $\sigma^y_i=i\sigma^x_i\sigma^z_i$, we obtain that $D_z$ transforms $\sigma^y_{i}\sigma^y_{i+1}\to
-\sigma^z_i\sigma^x_{i+1}\sigma^z_{i+2}$, or equivalently 
$H^{(-1)}\to-H^{(2)}$, and viceversa. By extending this argument to finite $m\neq0$, we find the relations summarized in Fig.~\ref{fig:scheme1}:
\begin{align}
D_z &: H_{0}^{(2m)} \to - H_{0}^{(1-2m)}\;,\label{eq:H0dualityX}\\
D_y &: H_{0}^{(-2m)} \to - H_{0}^{(-1+2m)}\;.\label{eq:H0dualityY}
\end{align}
This is the first main result of this paper: by a subsequent alternation of $D_z$ and $D_y$ we can generate all the $\mathbb{Z}$ phases of the non-interacting model. 

We now use the duality transformations to derive the order parameters of the different phases. Our starting point is the ferromagnetic model $H^{(1)}$, whose order parameter is $O^{(1)}\equiv\sigma^z$. By applying subsequently the duality transformations (\ref{eq:Dx}) and (\ref{eq:Dy}), we obtain the order parameters listed in the 3$^{\rm rd}$ column of Table \ref{table:order}. This list allows us to detect which symmetries are spontaneously broken in each phase (4$^{th}$ column). In general, we find that two phases break the same symmetry if and only if their $m$ indexes differ by a multiple of $4$. In particular, the phases $m=1$ and $m=-3$ are ferromagnets in the $z$ direction: both their order parameters break the symmetry $I_z:\sigma^z\to-\sigma^z$. 

The phases with even $m$'s have nonlocal (string) order parameter. String orders were extensively discussed in the context of topological phases of spins  \cite{den1989preroughening,kennedy1992hidden,oshikawa1992hidden,perez2008string,pollmann2010entanglement,pollmann2012detection}, bosons \cite{berg2008rise}, and fermions \cite{bahri2014detecting}. In particular, it was shown that string orders exist only in systems with unitary symmetries \cite{perez2008string}. Indeed, our model is symmetric under rotations of $\pi$ around the $x$ axis (which is equivalent to the subsequent application of the two Ising symmetries (\ref{eq:Ising})).


\begin{table*}[t]
\renewcommand{\arraystretch}{1.25}
\centering
\begin{tabular}{|x{0.5cm}|x{2.5cm}|x{2.5cm}|x{1.7cm}|x{1.7cm}|x{1.7cm}|}
\hline
$m$ & $H_0^{(m)}:$ parent Hamiltonian  & $O^{(m)}:$ order parameter  & broken symmetry & GS degeneracy & num. edge modes \\
\hline
4 & $\sigma^z\sigma^x\sigma^x\sigma^x\sigma^z $&  $\S^x\sigma^y\sigma^z\sigma^y\sigma^z$ & $\emptyset$ & 4 & 1 \\
\hline
3 & $\sigma^z\sigma^x\sigma^x\sigma^z $& $\sigma^z\sigma^y\sigma^z$ & $I_y$ & 8 & 1\\ 
\hline
2 & $\sigma^z\sigma^x\sigma^z $& $\S^x\sigma^y\sigma^z$ &  $\emptyset$ & 4 & 1\\
\hline
1 & $\sigma^z\sigma^z $& $\sigma^z$ & $I_z$ & 2 & 0\\
\hline
0 &$\sigma^x $&  $\S^x$ & $\emptyset$ & 1 & 0\\
\hline
-1 & $\sigma^y\sigma^y$& $\sigma^y$ & $I_y$ & 2 & 0\\
\hline
-2 & $\sigma^y\sigma^x\sigma^y $& $\S^x\sigma^z\sigma^y$ &  $\emptyset$ & 4& 1\\
\hline
-3 &$\sigma^y\sigma^x\sigma^x\sigma^y $&  $\sigma^y\sigma^z\sigma^y$ & $I_z$ & 8 & 1\\
\hline
\end{tabular}
\caption{Physical properties of 8 phases of spin chains with two Ising symmetries. In the 2$^{nd}$ and 3$^{rd}$ column, each operator acts on a neighboring site: for example $H_0^{(4)}=-\sum_i\sigma_{i}^z\sigma_{i+1}^x\sigma_{i+2}^x\sigma_{i+3}^x\sigma_{i+4}^z$ and $O^{(4)}_i=S^x_i\sigma^y_{i+1}\sigma^z_{i+2}\sigma^y_{i+3}\sigma^z_{i+4}$, where the string operator $\S^x_i$ is defined in Eq.~(\ref{eq:string}). The 4$^{\rm th}$ column indicates which symmetry is broken by the order parameter: for example the $m=3$ phase breaks the $I_y:\sigma^y\to-\sigma^y$ symmetry. The last two columns refer to non-integrable models within the same phase: the topologically non-trivial phases are characterized by one zero-energy mode per edge.}
\label{table:order}
\end{table*}

The order parameters $O^{(m)}$ can be used to rewrite the generalized cluster models (\ref{eq:H0spin}) in an appealing way:
\begin{align}
H_0^{(m)} = - \sum_i O^{(m)}_i O^{(m)}_{i+1}\;.\label{eq:H0order}
\end{align}
Because the order parameters $O^{(m)}$ at different sites commute \footnote{This can be verified either by a direct evaluation, or by noting that the order parameters at different sites are generated by the application of series of duality transformations to $\sigma^x$ operators belonging to different sites.},
$[O^{(m)}_i,O^{(m)}_j]=0$, these operators form a complete set of integrals of motion, with eigenvalues $\pm 1$ \footnote{Their eigenvalues can be determined by noting that $(O^{(m)}_i)^\dagger = O^{(m)}_i$ and $(O^{(m)}_i)^2 = 1$.}. Note that, in a system of size $L$, the Hamiltonian (\ref{eq:H0order}) has $L-|m|$ terms. This leads to a ground state degeneracy of $2^{|m|}$, in agreement with the fermionic analysis. This degeneracy has two distinct origins: symmetry breaking and edge states. For example, the ground state of $H^{(m=1)}$ is doubly degenerate due to the spontaneous breaking of the $I_z$ symmetry. The ground state of $H^{(m=-3)}$ has additionally two edge modes, $\sigma_1^y$ and $\sigma_L^y$, which commute with all $O^{(m)}_i$ and lead to a total degeneracy of $2^3=8$ \footnote{Note that the physics of edge modes is not captured by the duality transformations (\ref{eq:H0dualityX}) and (\ref{eq:H0dualityY}), which only apply to infinite systems with no boundaries.}. As expected, these edge states are invariant under the unbroken $I_y$ symmetry.

\emanuele{The duality transformations $D_z$ and $D_y$ are not only convenient theoretical tools to describe the SPT phases, but also have practical implications in the context of quantum computers. These many-body quantum systems are not characterized by an Hamiltonian, whose ground state can be reached by cooling to low temperatures. Instead, a quantum computer is initially prepared in a simple product state, usually $\Psi\rangle = \Pi_i |0\rangle_i$, on which one-qubit and two-qubit unitary operations can be applied. Ref.~\cite{choo2018measurement} demonstrated experimentally that using Hadamard (H) and controlled-Z (CZ) gates, one can prepare the ground state of the cluster Ising model, $m=2$. Their algorithm involves two steps: in the first step one applies H gates to each qubit independently and obtains the ground state of the $m=0$ Hamiltonian. Next, one applies CZ gates on all neighboring spins to obtain the ground state of the $m=2$ cluster Ising model. This second step is equivalent to the transformation $D_zD_y$, which indeed transforms the Hamiltonian $H_0^{(0)}$ to $H_0^{(2)}$ \cite{daniel}.}
	

\mysection{\label{interaction}Effects of integrability-breaking terms} The generalized cluster models of Eq.~(\ref{eq:H0}) have an infinite number of local order parameters ($O^{(m)}$ with odd $m$). However, when we consider the general class of models with two Ising symmetries, the number of order parameters must be reduced to two, one for each symmetry. To understand how this happens, we consider the effects of generic perturbations that preserve the two Ising symmetries (\ref{eq:Ising}). Under the Jordan-Wigner transformation, these terms are mapped to {\it local} interactions among the fermions: For example, $\sigma^x_i\sigma^x_{i+1}$ is mapped to the quartic term $\psi^\dagger_i\psi_i\psi^\dagger_{i+1}\psi_{i+1}$. In general, these fermionic terms preserve both $T$ (have a real coefficient) and $P$ (include an even number of fermionic operators). This type of interactions were shown to adiabatically  connect two phases if (and only if) their topological indexes differ by an integer multiple of 8. Thus, the number of topologically distinct phases is reduced to 8 \cite{fidkowski2010effects,fidkowski2011topological}. In the spin language, this leads to the 8 distinct phases listed in Table \ref{table:order}, and labeled by $m=-3,~...,~4$. The number and the symmetry of these phases coincide with the expectations based on the classification presented in Table \ref{table2}.


Interactions affect the order parameters as well and, in general,  couple all order parameters with the same symmetry. To exemplify this effect, let us consider the parent Hamiltonian $H^{(1)}$ whose ground state is a perfect ferromagnet in the $z$ direction $|\psi_{GS}\rangle=\prod_i|\sigma^z_i=+1\rangle$. According to {\em first order} perturbation theory, the effect of the interaction term $V_2=\sigma^x_i\sigma^x_{i+2}$ is proportional to $\langle O^{(-3)}_{i}\rangle \approx \langle \psi_{GS}|\sigma^y_{i}\sigma^z_{i+1}\sigma^y_{i+2}V_2|\psi_{GS}\rangle + c.c. \neq 0$.
This calculation shows that interaction can lead to a finite order parameter $O^{(-3)}$ in the phase $m=1$.

\begin{figure}
\centering
\includegraphics[scale=0.8]{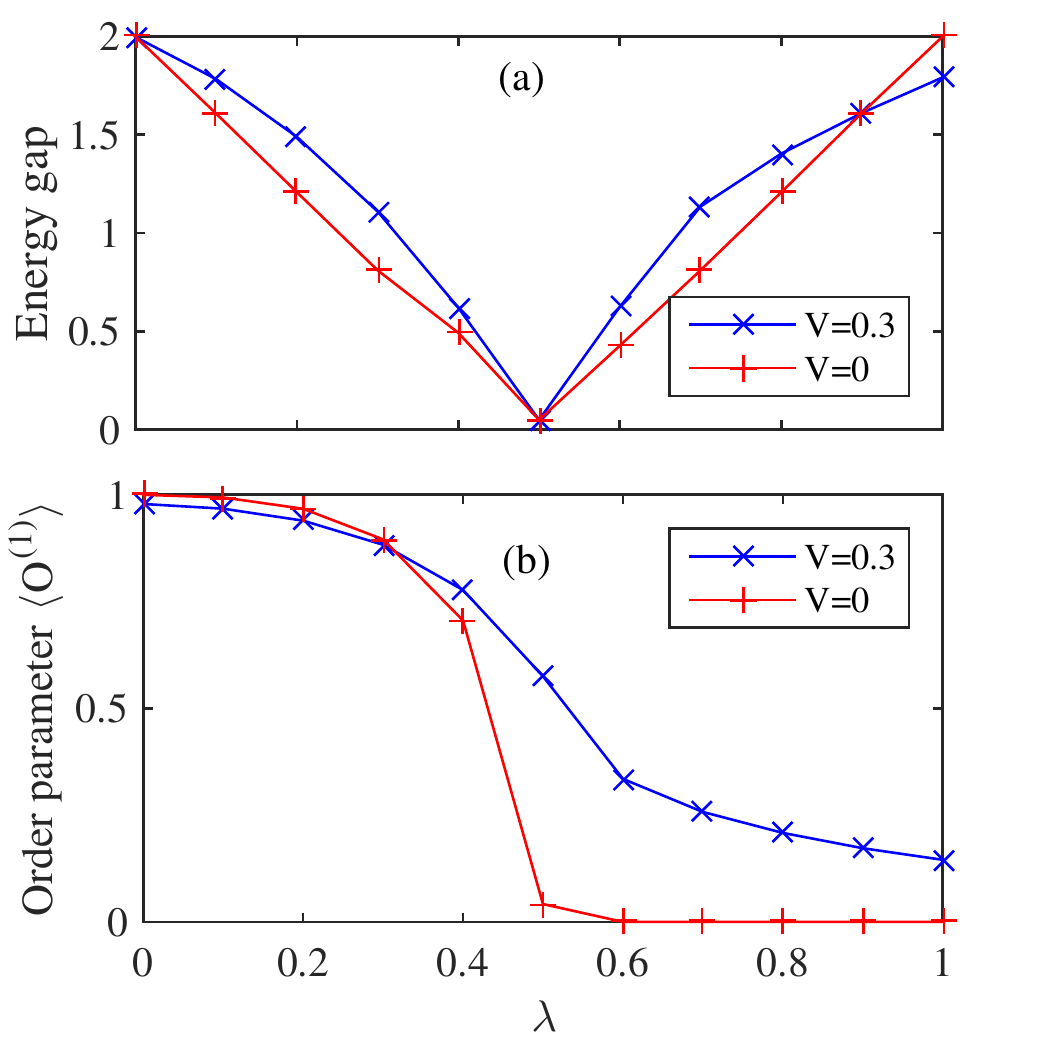}\\
\vspace{-0.3cm}
\caption{(a) Energy gap and (b) expectation value of the order parameter $O^{(1)} = \sigma^z$ across the transition $m=1\to m=-3$, for the integrable model ($V=0$), and in the presence of an integrability breaking term ($V=0.3$), see Eq.~(\ref{eq:Hij}). The phases $m=1$ and $m=-3$ break the symmetry $\sigma^z\to-\sigma^z$, but are topologically distinct thanks to the Ising symmetry $I_y:\sigma^y\to-\sigma^y$.}
\label{fig:numerics}
\end{figure}

To study the fate of the  order parameter $O^{(1)}=\sigma^z$ in the phase $m=-3$, we consider a model that interpolates between the parent Hamiltonians $ H_0^{(1)}$ and $H_0^{(-3)}$:
\begin{align}
&H(\lambda,V) = \sum_i ~(1-\lambda)~\sigma^z_i\sigma^z_{i+1} + ~\lambda~ \sigma^y_i\sigma^x_{i+1}\sigma^x_{i+2}\sigma^y_{i+3}\nonumber \\ &+ V \Big(\sigma^z_i\sigma^y_{i+1}\sigma^z_{i+2}\sigma^y_{i+3}+\sigma^y_i\sigma^z_{i+1}\sigma^y_{i+2}\sigma^z_{i+3}\Big)\;.
\label{eq:Hij}\end{align}
Here $V$ is a term that the breaks the integrability of the model, but preserves the two Ising symmetries. Fig.~\ref{fig:numerics} shows the numerically calculated excitation gap and the  order parameter $O^{(1)}$ as a function of $\lambda$ \footnote{The excitation gaps were computed by addressing the 8 lowest states and selecting the energy of the first non-degenerate excited state. Note that at the transition, the ground state degeneracy changes from 2 to 8. The order parameter is defined as the square-root of the ground-state correlation functions for distant spins. Fig.~\ref{fig:numerics} shows the result for a chain of size $L=128$. We verified that no significant finite size effect is observed, with the exception of the critical point $\lambda=0.5$. The calculations were performed using the ITensor C++ Library version 1.2, http://itensor.org.}. The key observation is that although the excitation gap (necessarily) closes when going from one phase to the other, the order parameter does not vanish at the transition. This finding shows that the phases $m=1$ and $m=-3$ are locally indistinguishable, and their distinction is of topological nature. Thanks to the duality transformations, we can extend this argument to any two phases with $m$ indexes that differ by a multiple of 4. We deduce that as expected, two phases share the same local order parameters, if and only if they break the same symmetry.

To determine the number of edge states in the SPT phases, we need to compare the degeneracy of the spin chains with open an closed boundary conditions. In the former case, the spin chains can be mapped to one dimensional fermionic models. In these models, interactions are known to lift some of the Majorana degeneracies, and in particular, change the ground state degeneracy of the fermionic model with 4 Majorana edge modes ($m=4$)  from $2^4=16$ to $4$ \cite{fidkowski2010effects,turner2011topological}. The degeneracy of the phases with $|m|\leq3$ are left unchanged, as summarized in Table \ref{table:order}. In the case of periodic boundary conditions, no edge states are present and the ground state degeneracy of the spin model is reduced to 1, for phases with no broken symmetry, or 2, if one Ising symmetry is broken. We conclude that in the topologically trivial phases ($m=0,~\pm 1$) the ground-state degeneracy of systems with open boundary conditions is equal to the case of periodic boundary condition. In contrast, in the topologically non-trivial phases ($|m|>1$), the former is 4 times larger than the latter. This observation is consistent with the presence of spin-1/2 edge modes in all three topologically non-trivial phases, $m=\pm 2,4$.

\mysection{General classification} We now connect the previous findings with the general classification of SPT phases of interacting bosons~\cite{chen2011complete,chen2013symmetry}. As mentioned in the introduction, the Ising symmetries (\ref{eq:Ising}) are antiunitary, and can be written as the product of a unitary transformation (a $\pi$ rotation around the $z$ or $y$ axis, respectively) and the spin time reversal symmetry ($\vec\sigma\to-\vec\sigma$). Indeed, under the Ising symmetry, the commutation relation of the spins operators changes sign and becomes $[\sigma^x,\sigma^y]= -2i\sigma^z$. Thus, the class of model under present consideration has two ${Z}^T_2$ symmetries, or equivalently a $Z^T_2\times Z_2$ symmetry

\emanuele{This observation shows how to extend our analysis to generic models of interacting bosons, given in terms of bosonic creation and anihilation operators, $b^\dagger$ and $b$. To express the two Ising symmetries in the bosonic language, we consider the approximate mapping $\sigma^x = 1-2 b^\dagger b$, $\sigma^y \approx i(b-b^\dagger)$, and $\sigma^z \approx b+b^\dagger$. This mapping becomes exact in the hard-core limit, where $b^\dagger b = 0,1$ and $\sigma^x = \pm 1$. The hard-core constraint is equivalent to the addition of a term $U(b^\dagger b-1/2)^2$, with $U\to\infty$, and thus, can be reached without affecting the symmetries of the model. Using this mapping we obtain that the two Ising symmetries map to the $P$ and $T$ bosonic symmetries, where $T$ is time-reversal symmetry (i.e. the requirement that the Hamiltonian can be written using creation and anihilation operators with real prefactors only) and  $P$ is parity (i.e. the conservation of the parity of the total number of bosons).}

\emanuele{As shown in Table \ref{table2}, one dimensional models with two Ising symmetries include 10 distinct phases (see Table \ref{table2}): When both symmetries are preserved, one obtains 4 topological phases \cite{chen2012symmetry,chen2013symmetry}. We identify these  phases as the  $m=0,\pm2,$ and $4$ phases of Table \ref{table:order}. An interesting question involves the identification of the Haldane phase of spin-1 chains, whose Heisenberg point belongs to our class of models. As shown by Ref.~\cite{fidkowski2011topological}, this model is topologically equivalent to 4 parallel Kitaev chains\footnote{Note that this result is in contradiction with an explicit statement of Ref.~\cite{verresen2017one}, who identified the Haldane phase with the cluster Ising model $H_0^{(2)}$. This point reserves further investigation.}, i.e. to the parent Hamiltonian $H_0^{(4)}$. This result can be rationalized as follows: the Heisenberg model is symmetric under any rotation, and in particular under the rotation of $\pi/2$ around the $x$ axis: This symmetry interchanges $z$ with $y$ and corresponds to $m\to-m$. Recalling that the phases $m$ and $m-8$ are topologically equivalent, we arrive to the conclusion that the Heisenberg model must belong to the $m=4$ phases, where $m-8=-4$.}
		
If one of the two Ising symmetries is broken, the problem is equivalent to the class of models with a single ${Z}_2^T$ symmetry, which has 2 distinct topological phases \cite{pollmann2010entanglement,pollmann2012symmetry,schuch2011classifying,chen2011classification,chen2011complete,pollmann2012detection}. This observation explains why our model has two distinct phases with ferromagnetic order in the $z$ direction, $m=1$ and $m=-3$ (2$^{nd}$ row of Table \ref{table2}). Because these two phases are topologically protected by an anti-unitary symmetry, they cannot be distinguished by both local and string order parameters \cite{perez2008string,pollmann2012symmetry}. A similar argument holds for the two ferromagnets in the $y$ direction (3$^{rd}$ row of Table \ref{table2}). These findings imply that the 8 phases listed in Table \ref{table:order} (corresponding to the 8 phases of interacting fermions with unbroken $P$ and $T$ symmetry) faithfully represent the first 3 rows of Table \ref{table2} (phases of spin chains with at least one unbroken Ising symmetry).

The remaining two phases of Table \ref{table2} can be constructed by considering fermionic models that spontaneously break T. For example, this occurs if the operator $O^T_i=i (\psi^\dagger_i+\psi_{i})(\psi^\dagger_{i+1}+\psi_{i+1})$ acquires a finite expectation value. The presence of the P symmetry places these models in the topological class $D$, which has 2 distinct topological phases. In the spin language, these two phases possess a finite expectation value of $O^T_i = \sigma^z_{i}\sigma^{y}_{i+1}$. This gives rise to two possible phases: a nematic phase with one order parameter $\langle\sigma^z_{i}\sigma^y_{i+1}\rangle\neq 0$, and a canted ferromagnet with two order parameters, $\langle\sigma^z_{i}\rangle \neq 0$ and $\langle\sigma^y_{i}\rangle \neq 0$. Note that the former phase has a residual $Z_2$ symmetry (rotations of $\pi$ around the $x$ axis), which is however insufficient to protect further topological phases.

\mysection{\label{sum}Summary and conclusions} In summary, we discussed the general phase diagram of spin chains with two Ising symmetries. This class of models is equivalent, under the Jordan-Wigner transformation, to 1d fermions with P and T symmetry. Accordingly, we find that this class of models has 10 distinct phases. The physical properties of the 10 phases of the fermions and of the spins are, however, very different. In particular, in the fermionic language 2 phases spontaneously break T, and the other 8 are of topological nature and do not break any symmetry. In contrast, in the spin language only 4 phases preserve both Ising symmetries, 4 phases break one of them, and 2 phases break both (see Table \ref{table2}).

Although we have not explicitly used this property, the parent Hamiltonians (\ref{eq:H0spin}) are translationally invariant. As shown in Refs.~\cite{turner2011topological,chen2011complete,chen2013symmetry}, this symmetry can double the number of distinct phases. In the case of an Ising coupling $\sigma^z_i\sigma^z_{i+1}$, for example, translationally invariant models can be classified as ferromagnets (if the prefactor is negative) or anti-ferromagnets (if the prefactor is positive). Similarly, the number of parent Hamiltonians (\ref{eq:H0spin}) can be doubled by considering positive prefactors. Possibly, this scenario can give rise to further SPT phases. Other interesting routes that are open by our work include 
the effects of periodic drives \cite{russomanno2016kibble,russomanno2017spin}, disorder \cite{strinati2017resilience,PhysRevB.65.104415}, or both \cite{else2016floquet,khemani2016phase}, as well as the extension of our study to 2 dimensions, where the classification of SPT phases predicts several yet-to-be-explored topologically non-trivial phases.

\acknowledgments We thank E. Berg, E. Demler, N. Lindner, D. Meidan, G. Murthy, M. Rudner, E. Shimshoni, and A. Soori  for many useful discussions. This work was supported by the Israeli Science Foundation Grant No. 1542/14. \emanuele{During the preparation of this manuscript we became aware of a related study now published in Ref.~\cite{verresen2017one}}



\bibliographystyle{naturemag}
\bibliography{KitaevChain}

\begin{thebibliography}{10}
\expandafter\ifx\csname url\endcsname\relax
  \def\url#1{\texttt{#1}}\fi
\expandafter\ifx\csname urlprefix\endcsname\relax\def\urlprefix{URL }\fi
\providecommand{\bibinfo}[2]{#2}
\providecommand{\eprint}[2][]{\url{#2}}

\bibitem{ando1975theory}
\bibinfo{author}{Ando, T.}, \bibinfo{author}{Matsumoto, Y.} \&
  \bibinfo{author}{Uemura, Y.}
\newblock \bibinfo{title}{Theory of hall effect in a two-dimensional electron
  system}.
\newblock \emph{\bibinfo{journal}{Journal of the Physical Society of Japan}}
  \textbf{\bibinfo{volume}{39}}, \bibinfo{pages}{279--288}
  (\bibinfo{year}{1975}).

\bibitem{klitzing1980new}
\bibinfo{author}{Klitzing, K.~v.}, \bibinfo{author}{Dorda, G.} \&
  \bibinfo{author}{Pepper, M.}
\newblock \bibinfo{title}{New method for high-accuracy determination of the
  fine-structure constant based on quantized hall resistance}.
\newblock \emph{\bibinfo{journal}{Physical Review Letters}}
  \textbf{\bibinfo{volume}{45}}, \bibinfo{pages}{494} (\bibinfo{year}{1980}).

\bibitem{laughlin1981quantized}
\bibinfo{author}{Laughlin, R.~B.}
\newblock \bibinfo{title}{Quantized hall conductivity in two dimensions}.
\newblock \emph{\bibinfo{journal}{Physical Review B}}
  \textbf{\bibinfo{volume}{23}}, \bibinfo{pages}{5632} (\bibinfo{year}{1981}).

\bibitem{halperin1982quantized}
\bibinfo{author}{Halperin, B.~I.}
\newblock \bibinfo{title}{Quantized hall conductance, current-carrying edge
  states, and the existence of extended states in a two-dimensional disordered
  potential}.
\newblock \emph{\bibinfo{journal}{Physical Review B}}
  \textbf{\bibinfo{volume}{25}}, \bibinfo{pages}{2185} (\bibinfo{year}{1982}).

\bibitem{kitaev2009periodic}
\bibinfo{author}{Kitaev, A.}
\newblock \bibinfo{title}{Periodic table for topological insulators and
  superconductors}.
\newblock In \emph{\bibinfo{booktitle}{AIP Conference Proceedings}}, vol.
  \bibinfo{volume}{1134 (1)}, \bibinfo{pages}{22--30}
  (\bibinfo{organization}{AIP}, \bibinfo{year}{2009}).

\bibitem{ryu2010topological}
\bibinfo{author}{Ryu, S.}, \bibinfo{author}{Schnyder, A.~P.},
  \bibinfo{author}{Furusaki, A.} \& \bibinfo{author}{Ludwig, A.~W.}
\newblock \bibinfo{title}{Topological insulators and superconductors: tenfold
  way and dimensional hierarchy}.
\newblock \emph{\bibinfo{journal}{New Journal of Physics}}
  \textbf{\bibinfo{volume}{12}}, \bibinfo{pages}{065010}
  (\bibinfo{year}{2010}).

\bibitem{hasan2010colloquium}
\bibinfo{author}{Hasan, M.~Z.} \& \bibinfo{author}{Kane, C.~L.}
\newblock \bibinfo{title}{Colloquium: topological insulators}.
\newblock \emph{\bibinfo{journal}{Reviews of Modern Physics}}
  \textbf{\bibinfo{volume}{82}}, \bibinfo{pages}{3045} (\bibinfo{year}{2010}).

\bibitem{haldane1983continuum}
\bibinfo{author}{Haldane, F. D.~M.}
\newblock \bibinfo{title}{Continuum dynamics of the 1-d heisenberg
  antiferromagnet: identification with the o (3) nonlinear sigma model}.
\newblock \emph{\bibinfo{journal}{Physics Letters A}}
  \textbf{\bibinfo{volume}{93}}, \bibinfo{pages}{464--468}
  (\bibinfo{year}{1983}).

\bibitem{haldane1983nonlinear}
\bibinfo{author}{Haldane, F.}
\newblock \bibinfo{title}{Nonlinear field theory of large-spin heisenberg
  antiferromagnets: semiclassically quantized solitons of the one-dimensional
  easy-axis n{\'e}el state}.
\newblock \emph{\bibinfo{journal}{Physical Review Letters}}
  \textbf{\bibinfo{volume}{50}}, \bibinfo{pages}{1153} (\bibinfo{year}{1983}).

\bibitem{affleck1987rigorous}
\bibinfo{author}{Affleck, I.}, \bibinfo{author}{Kennedy, T.},
  \bibinfo{author}{Lieb, E.~H.} \& \bibinfo{author}{Tasaki, H.}
\newblock \bibinfo{title}{Rigorous results on valence-bond ground states in
  antiferromagnets}.
\newblock \emph{\bibinfo{journal}{Physical Review Letters}}
  \textbf{\bibinfo{volume}{59}}, \bibinfo{pages}{799} (\bibinfo{year}{1987}).

\bibitem{dalla2006hidden}
\bibinfo{author}{Dalla~Torre, E.~G.}, \bibinfo{author}{Berg, E.} \&
  \bibinfo{author}{Altman, E.}
\newblock \bibinfo{title}{Hidden order in 1d bose insulators}.
\newblock \emph{\bibinfo{journal}{Physical Review Letters}}
  \textbf{\bibinfo{volume}{97}}, \bibinfo{pages}{260401}
  (\bibinfo{year}{2006}).

\bibitem{auerbach2012interacting}
\bibinfo{author}{Auerbach, A.}
\newblock \emph{\bibinfo{title}{Interacting electrons and quantum magnetism}}
  (\bibinfo{publisher}{Springer Science \& Business Media},
  \bibinfo{year}{2012}).

\bibitem{turner2013beyond}
\bibinfo{author}{Turner, A.~M.}, \bibinfo{author}{Vishwanath, A.} \&
  \bibinfo{author}{Head, C.~O.}
\newblock \bibinfo{title}{Beyond band insulators: topology of semimetals and
  interacting phases}.
\newblock \emph{\bibinfo{journal}{Topological Insulators}}
  \textbf{\bibinfo{volume}{6}}, \bibinfo{pages}{293--324}
  (\bibinfo{year}{2013}).

\bibitem{chen2013symmetry}
\bibinfo{author}{Chen, X.}, \bibinfo{author}{Gu, Z.-C.}, \bibinfo{author}{Liu,
  Z.-X.} \& \bibinfo{author}{Wen, X.-G.}
\newblock \bibinfo{title}{Symmetry protected topological orders and the group
  cohomology of their symmetry group}.
\newblock \emph{\bibinfo{journal}{Physical Review B}}
  \textbf{\bibinfo{volume}{87}}, \bibinfo{pages}{155114}
  (\bibinfo{year}{2013}).

\bibitem{senthil2015symmetry}
\bibinfo{author}{Senthil, T.}
\newblock \bibinfo{title}{Symmetry-protected topological phases of quantum
  matter}.
\newblock \emph{\bibinfo{journal}{Annu. Rev. Condens. Matter Phys.}}
  \textbf{\bibinfo{volume}{6}}, \bibinfo{pages}{299--324}
  (\bibinfo{year}{2015}).

\bibitem{chen2011classification}
\bibinfo{author}{Chen, X.}, \bibinfo{author}{Gu, Z.-C.} \&
  \bibinfo{author}{Wen, X.-G.}
\newblock \bibinfo{title}{Classification of gapped symmetric phases in
  one-dimensional spin systems}.
\newblock \emph{\bibinfo{journal}{Physical Review B}}
  \textbf{\bibinfo{volume}{83}}, \bibinfo{pages}{035107}
  (\bibinfo{year}{2011}).

\bibitem{chen2011complete}
\bibinfo{author}{Chen, X.}, \bibinfo{author}{Gu, Z.-C.} \&
  \bibinfo{author}{Wen, X.-G.}
\newblock \bibinfo{title}{Complete classification of one-dimensional gapped
  quantum phases in interacting spin systems}.
\newblock \emph{\bibinfo{journal}{Physical Review B}}
  \textbf{\bibinfo{volume}{84}}, \bibinfo{pages}{235128}
  (\bibinfo{year}{2011}).

\bibitem{chen2012symmetry}
\bibinfo{author}{Chen, X.}, \bibinfo{author}{Gu, Z.-C.}, \bibinfo{author}{Liu,
  Z.-X.} \& \bibinfo{author}{Wen, X.-G.}
\newblock \bibinfo{title}{Symmetry-protected topological orders in interacting
  bosonic systems}.
\newblock \emph{\bibinfo{journal}{Science}} \textbf{\bibinfo{volume}{338}},
  \bibinfo{pages}{1604--1606} (\bibinfo{year}{2012}).

\bibitem{berg2008rise}
\bibinfo{author}{Berg, E.}, \bibinfo{author}{Dalla~Torre, E.~G.},
  \bibinfo{author}{Giamarchi, T.} \& \bibinfo{author}{Altman, E.}
\newblock \bibinfo{title}{Rise and fall of hidden string order of lattice
  bosons}.
\newblock \emph{\bibinfo{journal}{Physical Review B}}
  \textbf{\bibinfo{volume}{77}}, \bibinfo{pages}{245119}
  (\bibinfo{year}{2008}).

\bibitem{pollmann2012symmetry}
\bibinfo{author}{Pollmann, F.}, \bibinfo{author}{Berg, E.},
  \bibinfo{author}{Turner, A.~M.} \& \bibinfo{author}{Oshikawa, M.}
\newblock \bibinfo{title}{Symmetry protection of topological phases in
  one-dimensional quantum spin systems}.
\newblock \emph{\bibinfo{journal}{Physical Review B}}
  \textbf{\bibinfo{volume}{85}}, \bibinfo{pages}{075125}
  (\bibinfo{year}{2012}).

\bibitem{pollmann2010entanglement}
\bibinfo{author}{Pollmann, F.}, \bibinfo{author}{Turner, A.~M.},
  \bibinfo{author}{Berg, E.} \& \bibinfo{author}{Oshikawa, M.}
\newblock \bibinfo{title}{Entanglement spectrum of a topological phase in one
  dimension}.
\newblock \emph{\bibinfo{journal}{Physical Review B}}
  \textbf{\bibinfo{volume}{81}}, \bibinfo{pages}{064439}
  (\bibinfo{year}{2010}).

\bibitem{schuch2011classifying}
\bibinfo{author}{Schuch, N.}, \bibinfo{author}{P{\'e}rez-Garc{\'\i}a, D.} \&
  \bibinfo{author}{Cirac, I.}
\newblock \bibinfo{title}{Classifying quantum phases using matrix product
  states and projected entangled pair states}.
\newblock \emph{\bibinfo{journal}{Physical Review B}}
  \textbf{\bibinfo{volume}{84}}, \bibinfo{pages}{165139}
  (\bibinfo{year}{2011}).

\bibitem{liu2011gapped}
\bibinfo{author}{Liu, Z.-X.}, \bibinfo{author}{Liu, M.} \&
  \bibinfo{author}{Wen, X.-G.}
\newblock \bibinfo{title}{Gapped quantum phases for the s= 1 spin chain with
  $d_{2h}$ symmetry}.
\newblock \emph{\bibinfo{journal}{Physical Review B}}
  \textbf{\bibinfo{volume}{84}}, \bibinfo{pages}{075135}
  (\bibinfo{year}{2011}).

\bibitem{jordan1928paulische}
\bibinfo{author}{Jordan, P.} \& \bibinfo{author}{Wigner, E.}
\newblock \bibinfo{title}{{\"u}ber das paulische {\"a}quivalenzverbot}.
\newblock \emph{\bibinfo{journal}{Zeitschrift f{\"u}r Physik A Hadrons and
  Nuclei}} \textbf{\bibinfo{volume}{47}}, \bibinfo{pages}{631--651}
  (\bibinfo{year}{1928}).

\bibitem{sachdev2007quantum}
\bibinfo{author}{Sachdev, S.}
\newblock \emph{\bibinfo{title}{Quantum phase transitions}}
  (\bibinfo{publisher}{Wiley Online Library}, \bibinfo{year}{2007}).

\bibitem{kitaev2001unpaired}
\bibinfo{author}{Kitaev, A.~Y.}
\newblock \bibinfo{title}{Unpaired majorana fermions in quantum wires}.
\newblock \emph{\bibinfo{journal}{Physics-Uspekhi}}
  \textbf{\bibinfo{volume}{44}}, \bibinfo{pages}{131} (\bibinfo{year}{2001}).

\bibitem{greiter20141d}
\bibinfo{author}{Greiter, M.}, \bibinfo{author}{Schnells, V.} \&
  \bibinfo{author}{Thomale, R.}
\newblock \bibinfo{title}{The 1d ising model and the topological phase of the
  kitaev chain}.
\newblock \emph{\bibinfo{journal}{Annals of Physics}}
  \textbf{\bibinfo{volume}{351}}, \bibinfo{pages}{1026--1033}
  (\bibinfo{year}{2014}).

\bibitem{alicea2010non}
\bibinfo{author}{Alicea, J.}, \bibinfo{author}{Oreg, Y.},
  \bibinfo{author}{Refael, G.}, \bibinfo{author}{Von~Oppen, F.} \&
  \bibinfo{author}{Fisher, M.~P.}
\newblock \bibinfo{title}{Non-abelian statistics and topological quantum
  information processing in 1d wire networks}.
\newblock \emph{\bibinfo{journal}{Nature Physics}}
  \textbf{\bibinfo{volume}{7}}, \bibinfo{pages}{412} (\bibinfo{year}{2011}).

\bibitem{fidkowski2011topological}
\bibinfo{author}{Fidkowski, L.} \& \bibinfo{author}{Kitaev, A.}
\newblock \bibinfo{title}{Topological phases of fermions in one dimension}.
\newblock \emph{\bibinfo{journal}{Physical Review B}}
  \textbf{\bibinfo{volume}{83}}, \bibinfo{pages}{075103}
  (\bibinfo{year}{2011}).

\bibitem{niu2012majorana}
\bibinfo{author}{Niu, Y.} \emph{et~al.}
\newblock \bibinfo{title}{Majorana zero modes in a quantum ising chain with
  longer-ranged interactions}.
\newblock \emph{\bibinfo{journal}{Physical Review B}}
  \textbf{\bibinfo{volume}{85}}, \bibinfo{pages}{035110}
  (\bibinfo{year}{2012}).

\bibitem{zirnbauer1996riemannian}
\bibinfo{author}{Zirnbauer, M.~R.}
\newblock \bibinfo{title}{Riemannian symmetric superspaces and their origin in
  random-matrix theory}.
\newblock \emph{\bibinfo{journal}{Journal of Mathematical Physics}}
  \textbf{\bibinfo{volume}{37}}, \bibinfo{pages}{4986--5018}
  (\bibinfo{year}{1996}).

\bibitem{altland1997nonstandard}
\bibinfo{author}{Altland, A.} \& \bibinfo{author}{Zirnbauer, M.~R.}
\newblock \bibinfo{title}{Nonstandard symmetry classes in mesoscopic
  normal-superconducting hybrid structures}.
\newblock \emph{\bibinfo{journal}{Physical Review B}}
  \textbf{\bibinfo{volume}{55}}, \bibinfo{pages}{1142} (\bibinfo{year}{1997}).

\bibitem{suzuki1971relationship}
\bibinfo{author}{Suzuki, M.}
\newblock \bibinfo{title}{Relationship among exactly soluble models of critical
  phenomena. i) 2d ising model, dimer problem and the generalized
  {{XY}}-model}.
\newblock \emph{\bibinfo{journal}{Progress of Theoretical Physics}}
  \textbf{\bibinfo{volume}{46}}, \bibinfo{pages}{1337--1359}
  (\bibinfo{year}{1971}).

\bibitem{keating2004random}
\bibinfo{author}{Keating, J.} \& \bibinfo{author}{Mezzadri, F.}
\newblock \bibinfo{title}{Random matrix theory and entanglement in quantum spin
  chains}.
\newblock \emph{\bibinfo{journal}{Communications in mathematical physics}}
  \textbf{\bibinfo{volume}{252}}, \bibinfo{pages}{543--579}
  (\bibinfo{year}{2004}).

\bibitem{kopp2005criticality}
\bibinfo{author}{Kopp, A.} \& \bibinfo{author}{Chakravarty, S.}
\newblock \bibinfo{title}{Criticality in correlated quantum matter}.
\newblock \emph{\bibinfo{journal}{Nature Physics}}
  \textbf{\bibinfo{volume}{1}}, \bibinfo{pages}{53--56} (\bibinfo{year}{2005}).

\bibitem{son2011quantum}
\bibinfo{author}{Son, W.} \emph{et~al.}
\newblock \bibinfo{title}{Quantum phase transition between cluster and
  antiferromagnetic states}.
\newblock \emph{\bibinfo{journal}{EPL (Europhysics Letters)}}
  \textbf{\bibinfo{volume}{95}}, \bibinfo{pages}{50001} (\bibinfo{year}{2011}).

\bibitem{smacchia2011statistical}
\bibinfo{author}{Smacchia, P.} \emph{et~al.}
\newblock \bibinfo{title}{Statistical mechanics of the cluster ising model}.
\newblock \emph{\bibinfo{journal}{Physical Review A}}
  \textbf{\bibinfo{volume}{84}}, \bibinfo{pages}{022304}
  (\bibinfo{year}{2011}).

\bibitem{degottardi2013majorana}
\bibinfo{author}{DeGottardi, W.}, \bibinfo{author}{Thakurathi, M.},
  \bibinfo{author}{Vishveshwara, S.} \& \bibinfo{author}{Sen, D.}
\newblock \bibinfo{title}{Majorana fermions in superconducting wires: Effects
  of long-range hopping, broken time-reversal symmetry, and potential
  landscapes}.
\newblock \emph{\bibinfo{journal}{Physical Review B}}
  \textbf{\bibinfo{volume}{88}}, \bibinfo{pages}{165111}
  (\bibinfo{year}{2013}).

\bibitem{lahtinen2015realizing}
\bibinfo{author}{Lahtinen, V.} \& \bibinfo{author}{Ardonne, E.}
\newblock \bibinfo{title}{Realizing all $s o (n)_ 1$ quantum criticalities in
  symmetry protected cluster models}.
\newblock \emph{\bibinfo{journal}{Physical Review Letters}}
  \textbf{\bibinfo{volume}{115}}, \bibinfo{pages}{237203}
  (\bibinfo{year}{2015}).

\bibitem{ohta2016topological}
\bibinfo{author}{Ohta, T.}, \bibinfo{author}{Tanaka, S.},
  \bibinfo{author}{Danshita, I.} \& \bibinfo{author}{Totsuka, K.}
\newblock \bibinfo{title}{Topological and dynamical properties of a generalized
  cluster model in one dimension}.
\newblock \emph{\bibinfo{journal}{Physical Review B}}
  \textbf{\bibinfo{volume}{93}}, \bibinfo{pages}{165423}
  (\bibinfo{year}{2016}).

\bibitem{lee2016string}
\bibinfo{author}{Lee, T.~E.}, \bibinfo{author}{Joglekar, Y.~N.} \&
  \bibinfo{author}{Richerme, P.}
\newblock \bibinfo{title}{String order via floquet interactions in atomic
  systems}.
\newblock \emph{\bibinfo{journal}{Physical Review A}}
  \textbf{\bibinfo{volume}{94}}, \bibinfo{pages}{023610}
  (\bibinfo{year}{2016}).

\bibitem{russomanno2017spin}
\bibinfo{author}{Russomanno, A.}, \bibinfo{author}{Dalla~Torre, E.~G.}
  \emph{et~al.}
\newblock \bibinfo{title}{Spin and topological order in a periodically driven
  spin chain}.
\newblock \emph{\bibinfo{journal}{Physical Review B}}
  \textbf{\bibinfo{volume}{96}}, \bibinfo{pages}{045422}
  (\bibinfo{year}{2017}).

\bibitem{briegel2001persistent}
\bibinfo{author}{Briegel, H.~J.} \& \bibinfo{author}{Raussendorf, R.}
\newblock \bibinfo{title}{Persistent entanglement in arrays of interacting
  particles}.
\newblock \emph{\bibinfo{journal}{Physical Review Letters}}
  \textbf{\bibinfo{volume}{86}}, \bibinfo{pages}{910} (\bibinfo{year}{2001}).

\bibitem{raussendorf2001one}
\bibinfo{author}{Raussendorf, R.} \& \bibinfo{author}{Briegel, H.~J.}
\newblock \bibinfo{title}{A one-way quantum computer}.
\newblock \emph{\bibinfo{journal}{Physical Review Letters}}
  \textbf{\bibinfo{volume}{86}}, \bibinfo{pages}{5188--5191}
  (\bibinfo{year}{2001}).

\bibitem{doherty2009identifying}
\bibinfo{author}{Doherty, A.~C.} \& \bibinfo{author}{Bartlett, S.~D.}
\newblock \bibinfo{title}{Identifying phases of quantum many-body systems that
  are universal for quantum computation}.
\newblock \emph{\bibinfo{journal}{Physical Review Letters}}
  \textbf{\bibinfo{volume}{103}}, \bibinfo{pages}{020506}
  (\bibinfo{year}{2009}).

\bibitem{stephen2017computational}
\bibinfo{author}{Stephen, D.~T.}, \bibinfo{author}{Wang, D.-S.},
  \bibinfo{author}{Prakash, A.}, \bibinfo{author}{Wei, T.-C.} \&
  \bibinfo{author}{Raussendorf, R.}
\newblock \bibinfo{title}{Computational power of symmetry-protected topological
  phases}.
\newblock \emph{\bibinfo{journal}{Physical Review Letters}}
  \textbf{\bibinfo{volume}{119}}, \bibinfo{pages}{010504}
  (\bibinfo{year}{2017}).

\bibitem{den1989preroughening}
\bibinfo{author}{den Nijs, M.} \& \bibinfo{author}{Rommelse, K.}
\newblock \bibinfo{title}{Preroughening transitions in crystal surfaces and
  valence-bond phases in quantum spin chains}.
\newblock \emph{\bibinfo{journal}{Physical Review B}}
  \textbf{\bibinfo{volume}{40}}, \bibinfo{pages}{4709} (\bibinfo{year}{1989}).

\bibitem{kennedy1992hidden}
\bibinfo{author}{Kennedy, T.} \& \bibinfo{author}{Tasaki, H.}
\newblock \bibinfo{title}{Hidden $z_2\times z_2$ symmetry breaking in
  haldane-gap antiferromagnets}.
\newblock \emph{\bibinfo{journal}{Physical Review B}}
  \textbf{\bibinfo{volume}{45}}, \bibinfo{pages}{304} (\bibinfo{year}{1992}).

\bibitem{oshikawa1992hidden}
\bibinfo{author}{Oshikawa, M.}
\newblock \bibinfo{title}{Hidden $z_2* z_2$ symmetry in quantum spin chains
  with arbitrary integer spin}.
\newblock \emph{\bibinfo{journal}{Journal of Physics: Condensed Matter}}
  \textbf{\bibinfo{volume}{4}}, \bibinfo{pages}{7469} (\bibinfo{year}{1992}).

\bibitem{perez2008string}
\bibinfo{author}{P{\'e}rez-Garc{\'\i}a, D.}, \bibinfo{author}{Wolf, M.~M.},
  \bibinfo{author}{Sanz, M.}, \bibinfo{author}{Verstraete, F.} \&
  \bibinfo{author}{Cirac, J.~I.}
\newblock \bibinfo{title}{String order and symmetries in quantum spin
  lattices}.
\newblock \emph{\bibinfo{journal}{Physical Review Letters}}
  \textbf{\bibinfo{volume}{100}}, \bibinfo{pages}{167202}
  (\bibinfo{year}{2008}).

\bibitem{pollmann2012detection}
\bibinfo{author}{Pollmann, F.} \& \bibinfo{author}{Turner, A.~M.}
\newblock \bibinfo{title}{Detection of symmetry-protected topological phases in
  one dimension}.
\newblock \emph{\bibinfo{journal}{Physical Review B}}
  \textbf{\bibinfo{volume}{86}}, \bibinfo{pages}{125441}
  (\bibinfo{year}{2012}).

\bibitem{bahri2014detecting}
\bibinfo{author}{Bahri, Y.} \& \bibinfo{author}{Vishwanath, A.}
\newblock \bibinfo{title}{Detecting majorana fermions in quasi-one-dimensional
  topological phases using nonlocal order parameters}.
\newblock \emph{\bibinfo{journal}{Physical Review B}}
  \textbf{\bibinfo{volume}{89}}, \bibinfo{pages}{155135}
  (\bibinfo{year}{2014}).

\bibitem{choo2018measurement}
\bibinfo{author}{Choo, K.}, \bibinfo{author}{von Keyserlingk, C.~W.},
  \bibinfo{author}{Regnault, N.} \& \bibinfo{author}{Neupert, T.}
\newblock \bibinfo{title}{Measurement of the entanglement spectrum of a
  symmetry-protected topological state using the {{IBM}} quantum computer}.
\newblock \emph{\bibinfo{journal}{Physical Review Letters}}
  \textbf{\bibinfo{volume}{121}}, \bibinfo{pages}{086808}
  (\bibinfo{year}{2018}).

\bibitem{daniel}
\bibinfo{author}{Atzitz, D.}, \bibinfo{author}{Sela, E.} \&
  \bibinfo{author}{Dalla~Torre, E.~G.}
\newblock \bibinfo{title}{Entanglement entropy and negativity of generalized
  cluster ising models}.
\newblock \emph{\bibinfo{journal}{in preparation}}  (\bibinfo{year}{2018}).

\bibitem{fidkowski2010effects}
\bibinfo{author}{Fidkowski, L.} \& \bibinfo{author}{Kitaev, A.}
\newblock \bibinfo{title}{Effects of interactions on the topological
  classification of free fermion systems}.
\newblock \emph{\bibinfo{journal}{Physical Review B}}
  \textbf{\bibinfo{volume}{81}}, \bibinfo{pages}{134509}
  (\bibinfo{year}{2010}).

\bibitem{turner2011topological}
\bibinfo{author}{Turner, A.~M.}, \bibinfo{author}{Pollmann, F.} \&
  \bibinfo{author}{Berg, E.}
\newblock \bibinfo{title}{Topological phases of one-dimensional fermions: An
  entanglement point of view}.
\newblock \emph{\bibinfo{journal}{Physical Review B}}
  \textbf{\bibinfo{volume}{83}}, \bibinfo{pages}{075102}
  (\bibinfo{year}{2011}).

\bibitem{verresen2017one}
\bibinfo{author}{Verresen, R.}, \bibinfo{author}{Moessner, R.} \&
  \bibinfo{author}{Pollmann, F.}
\newblock \bibinfo{title}{One-dimensional symmetry protected topological phases
  and their transitions}.
\newblock \emph{\bibinfo{journal}{Physical Review B}}
  \textbf{\bibinfo{volume}{96}}, \bibinfo{pages}{165124}
  (\bibinfo{year}{2017}).

\bibitem{russomanno2016kibble}
\bibinfo{author}{Russomanno, A.} \& \bibinfo{author}{Dalla~Torre, E.~G.}
\newblock \bibinfo{title}{Kibble-zurek scaling in periodically driven quantum
  systems}.
\newblock \emph{\bibinfo{journal}{EPL (Europhysics Letters)}}
  \textbf{\bibinfo{volume}{115}}, \bibinfo{pages}{30006}
  (\bibinfo{year}{2016}).

\bibitem{strinati2017resilience}
\bibinfo{author}{Strinati, M.~C.}, \bibinfo{author}{Rossini, D.},
  \bibinfo{author}{Fazio, R.} \& \bibinfo{author}{Russomanno, A.}
\newblock \bibinfo{title}{Resilience of hidden order to symmetry-preserving
  disorder}.
\newblock \emph{\bibinfo{journal}{Physical Review B}}
  \textbf{\bibinfo{volume}{96}}, \bibinfo{pages}{214206}
  (\bibinfo{year}{2017}).

\bibitem{PhysRevB.65.104415}
\bibinfo{author}{M\'elin, R.}, \bibinfo{author}{Lin, Y.-C.},
  \bibinfo{author}{Lajk\'o, P.}, \bibinfo{author}{Rieger, H.} \&
  \bibinfo{author}{Igl\'oi, F.}
\newblock \bibinfo{title}{Strongly disordered spin ladders}.
\newblock \emph{\bibinfo{journal}{Physical Review B}}
  \textbf{\bibinfo{volume}{65}}, \bibinfo{pages}{104415}
  (\bibinfo{year}{2002}).

\bibitem{else2016floquet}
\bibinfo{author}{Else, D.~V.}, \bibinfo{author}{Bauer, B.} \&
  \bibinfo{author}{Nayak, C.}
\newblock \bibinfo{title}{Floquet time crystals}.
\newblock \emph{\bibinfo{journal}{Physical Review Letters}}
  \textbf{\bibinfo{volume}{117}}, \bibinfo{pages}{090402}
  (\bibinfo{year}{2016}).

\bibitem{khemani2016phase}
\bibinfo{author}{Khemani, V.}, \bibinfo{author}{Lazarides, A.},
  \bibinfo{author}{Moessner, R.} \& \bibinfo{author}{Sondhi, S.~L.}
\newblock \bibinfo{title}{Phase structure of driven quantum systems}.
\newblock \emph{\bibinfo{journal}{Physical Review Letters}}
  \textbf{\bibinfo{volume}{116}}, \bibinfo{pages}{250401}
  (\bibinfo{year}{2016}).

\end{thebibliography}

\newpage

\end{document}